\pdfoutput=1

\documentclass[11pt]{article}

\usepackage[final]{acl}

\usepackage{times}
\usepackage{latexsym}

\usepackage[T1]{fontenc}

\usepackage[utf8]{inputenc}

\usepackage{microtype}

\usepackage{inconsolata}

\usepackage{graphicx}

\usepackage[acronym]{glossaries}
\title{Speech Editing -- a Summary}

\author{Tobias K\"assmann \\
  \texttt{tobias.kaessmann@kit.edu} \\\And
  Yining Liu \\
  \texttt{yining.liu9@kit.edu} \\\And
  Danni Liu\\
  danni.liu@kit.edu}

\begin{document}
\maketitle
\begin{abstract}
With the rise of video production and social media, speech editing has become crucial for creators to address issues like mispronunciations, missing words, or stuttering in audio recordings. This paper explores text-based speech editing methods that modify audio via text transcripts without manual waveform editing. These approaches ensure edited audio is indistinguishable from the original by altering the mel-spectrogram. Recent advancements, such as context-aware prosody correction and advanced attention mechanisms, have improved speech editing quality. This paper reviews state-of-the-art methods, compares key metrics, and examines widely used datasets. The aim is to highlight ongoing issues and inspire further research and innovation in speech editing.
\end{abstract}

\newacronym{mos}{MOS}{Mean Opinion Score}
\newacronym{mcd}{MCD}{Mel-Cepstral distance measure}
\newacronym{wer}{WER}{Word Error Rate}
\newacronym{sim}{SIM}{Speaker Similarity}
\newacronym{dtw}{DTW}{Dynamic Time Warping}

\newacronym{tts}{TTS}{Text to speech}

\newacronym{smos}{sMOS}{similarity MOS}
\newacronym{qmos}{qMOS}{quality MOS}
\newacronym{cmos}{cMOS}{comparative mean opinion score}
\newacronym{DNSMos}{DNSMos}{Dynamic Time Warping}

\newacronym{itu}{ITU}{International Telecommunication Union}
\newacronym{mushra}{MUSHRA}{Multiple Stimuli with Hidden Reference and Anchor}

\newacronym{pesq}{PESQ}{Perceptual evaluation of speech quality}
\newacronym{stoi}{STOI}{short-time objective intelligibility measure}

\newacronym{nn}{NN}{Neural Network}
\newacronym{hmm}{HMM}{Hidden Markov Model}
\newacronym{g2p}{G2P}{Grapheme-to-Phoneme}
\newacronym{gan}{GAN}{Generative Adversarial Network}
\newacronym{lstm}{LSTM}{Long short-term memory}
\newacronym{ar}{AR}{Autoregressive}
\newacronym{cnn}{CNN}{Convolutional Neural Network}
\newacronym{ntts}{NTTS}{Neural Text-to-Speech}

\section{Introduction}
As audio data becomes increasingly prevalent in our daily lives, especially due to the rise of video production and audio sharing on social media platforms like Instagram, speech editing has become a crucial tool for improving creators' efficiency. Issues such as mispronunciations, missing words, or stuttering can disrupt recordings, making speech editing essential to avoid the need for redoing entire recordings \cite{SpeechX, EditSpeech}. A text-based speech editing approach is particularly beneficial for post-editing audio data using a text transcript \cite{Morrison2021}.

\begin{figure}
    \centering
    \includegraphics[width=\linewidth]{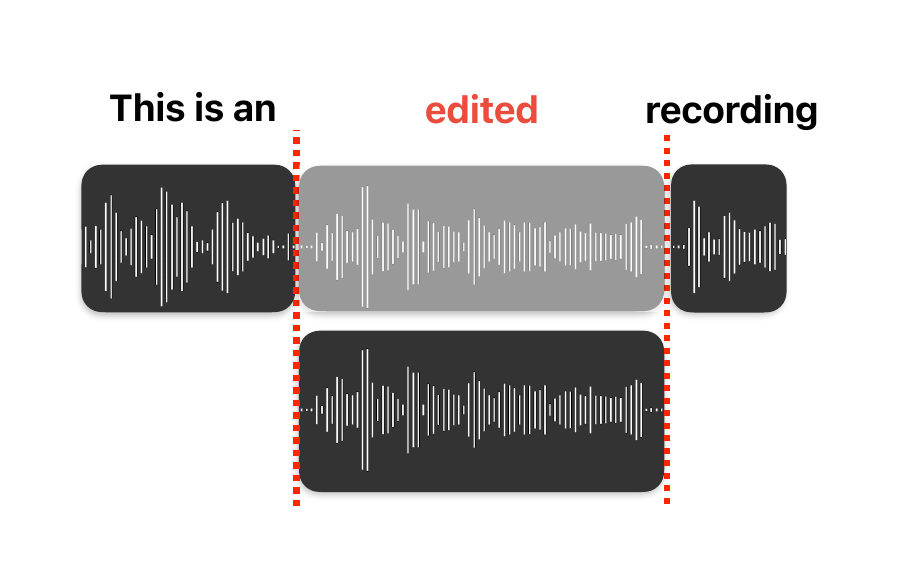}
    \caption{Speech Editing by inpainting (Example Illustration of an additional word getting added that was not present before)}
    \label{fig:Text_edit_example}
\end{figure}

Speech editing involves altering words and phrases based on a text transcript of an audio utterance. This can be achieved by textually deleting or cutting unwanted words and phrases, copying and pasting them to different locations within the text, or even inserting new unseen text, which has to be generated for this purpose into the text-transcript, as illustrated in Figure \ref{fig:Text_edit_example}. These changes are then propagated to the audio recording, by altering the mel-spectrogram \cite{mel-spectogram} of the original audio, without the need for manual waveform editing \cite{Morrison2021, SpeechX, RetrieverTTS}. The goal is to produce edited audio that is indistinguishable from the original, while ensuring that the original audio fragments remain unchanged \cite{AttentionStitch, VoiceCraft}.

One major challenge in speech editing is ensuring that the boundaries of the edited regions are seamlessly integrated in terms of prosody\footnote{the patterns of stress and intonation in a language}. This can be difficult when inserted words or phrases do not match the surrounding context in terms of intonation, stress, rhythm, voice quality, similarity, temporal smoothness, style, or naturalness \cite{Morrison2021, SpeechX, Mapache, EditSpeech, Editts}. Additionally, external factors such as environmental noise or packet loss can affect the quality of the edited audio \cite{SpeechPainter}.

Neglected at first, recent methods have also emphasized the importance of stutter removal in speech editing \cite{FluentSpeech}. Existing methods often struggle with robustness when dealing with stuttered speech, leading to over-smoothed outputs that fail to meet quality criteria and do not generalize well to unseen speakers \cite{SpeechPainter, Mapache, FluentSpeech}. Addressing these challenges is essential for developing effective speech editing tools that enhance the quality and naturalness of edited audio.

In order to help with this task, our goal with this paper is to provide a summary of the topic, a comparison of some noteworthy papers, and further insights into the current state of the art. We aim to draw more attention to this research area, encourage more researchers to engage with this topic, and highlight current issues.

\section{Metrics}
Currently, there are many methods used to determine the performance of a speech edit system. The most noteworthy ones include the \gls{mos}, \gls{mcd}, \gls{wer}, \gls{sim} because they are the most commonly used metrics as can be seen in Table \ref{table:objective_models_comparison}, that presents a comparison of objective measures, and Table \ref{table:subjective_models_comparison} which provides a comparison of subjective measures used in various papers and their results.

\subsection{MOS}
\label{mos-section}
\gls{mos} is the most common measure within the speech editing community, as almost every research paper uses it or a variant thereof. \gls{mos} is a score that subjectively measures audio quality based on ratings by human listeners \cite{AttentionStitch}. The \gls{mos} score is calculated by asking human evaluators to rate the audio based on quality criteria on a $1$ to $5$ Likert scale, where $1$ means very unrealistic and $5$ means very realistic speech \cite{VoiceBox, EditSpeech, RetrieverTTS, SpeechPainter}. According to our own definition, we hereby declare that "realistic" in this context means that the speech is coherent with the context, has the correct textual content, and sounds natural, including pronunciation, intonation, naturalness, and clarity. The arithmetic mean of these ratings is then calculated, with a higher \gls{mos} value indicating better quality. It is important to note that \gls{mos} is also a recomme ndation by the \gls{itu}\footnote{https://www.itu.int/rec/T-REC-P.85/en  -- ITU-T Recommendation P.85, 1994. Telephone transmission quality subjective opinion tests. A method for subjective performance assessment of the quality of speech voice output devices}.

To gather sufficient data, some research papers use Amazon Mechanical Turk \cite{AmazonMechanicalTurk} like \cite{A3T, Morrison2021, FluentSpeech}. In the case of \cite{Morrison2021}, even a listening test was conducted to assess the hearing capabilities of the testers and to prevent wrong test results.

An issue with the \gls{mos} score is that it is not comparable across different papers \cite{VISWANATHAN200555, VoiceBox}, leading to the development of various \gls{mos} variants, such as \gls{smos} \cite{VoiceBox, RetrieverTTS}, \gls{qmos} \cite{VoiceBox}, \gls{cmos} \cite{FluentSpeech}, \gls{DNSMos} \cite{DNSMOS, SpeechX} or one with a modified scale based on psychometric properties \cite{VISWANATHAN200555}. For instance, \cite{FluentSpeech} employs \gls{cmos}, a comparative mean opinion score. \cite{SpeechX} uses \gls{DNSMos} to evaluate noise suppression and target speaker performance, utilizing the overall score from the \gls{DNSMos}\footnote{DNSMOS P.835} model and a personalized \gls{DNSMos} model tailored for target speaker performance evaluation.

The similarity \gls{mos} (\gls{smos}) used by \cite{VoiceBox, RetrieverTTS} measures subjective audio similarity given pairs of audio, while the \gls{qmos} employed by \cite{VoiceBox} measures subjective quality.

The sheer number of \gls{mos} variants and the subjective nature of \gls{mos} call for an automatic measure like MosNet \cite{MosNet}. However, apart from \cite{SpeechX} with the use of \gls{DNSMos} \cite{DNSMOS, DNSMOS835}, automatic measures are not widely used in the speech community, likely due to the complexity of evaluating human voice and the challenges models like MosNet \cite{MosNet} still face in evaluation.

\subsection{MCD}

Another rather popular measure is the \gls{mcd} \cite{Kubichek1993} measure, a common objective metric used by several studies \cite{AttentionStitch, A3T, EditSpeech, FluentSpeech, SpeechX, Jin2017}. It computes the spectral dissimilarity between two mel-spectrograms \cite{AttentionStitch, mel-spectogram}, where a lower \gls{mcd} value indicates better speech quality \cite{EditSpeech}.

The \gls{mcd} analysis can be performed on three segments: the modified region, the unmodified region, and the entire utterance. However, aside from \cite{EditSpeech}, no other study we found has investigated these segments in detail. Most studies compare the synthesized utterance directly to the reference (altered) audio.

Although the \gls{mcd} is widely used in the voice conversion (VC) community to automatically measure the quality of converted speech, it has its limitations. One significant drawback is the lack of correlation with human perception, as the function only measures the distortion of acoustic features and does not capture subjective human judgments \cite{MosNet}. Another limitation is that it assumes a deterministic output given the input, which is often unrealistic. Therefore, \cite{VoiceBox} advocates for other reproducible model-based perceptual metrics.

\subsection{SIM and WER}
In addition to the objective \gls{mcd} measure, other objective measures such as the \gls{wer} and \gls{sim} are often used within the community to provide additional means of comparing their works.

\gls{sim} measures the coherence of audio utterances. This is done by calculating the cosine similarity difference between the generated embeddings and the ground truth embeddings of the audio utterances \cite{Mapache, VoiceBox, vall-E, SpeechX, VoiceCraft, Kharitonov2021}. There are several ways to generate these embeddings, such as the WavLM-TDCNN embeddings \cite{Chen2022} used by \cite{VoiceCraft, Mapache}, NeMo’s TitaNet-Large \cite{SpeechX}, or the Google Cloud speech-to-text API \cite{Editts}.

The Google Cloud speech-to-text API\footnote{\url{https://cloud.google.com/speech-to-text/?hl=en}} was also used by \cite{Editts} to examine the \gls{wer}, which measures the fidelity of the generated audio with respect to its provided transcription \cite{SpeechX}. Other papers measured \gls{wer} using the open-sourced Whisper-Medium ASR \cite{Mapache, Radford2022}, HuBERT \cite{Hsu2021} for English settings. For multilingual settings also Whisper \cite{Radford2023} can be used as in the works of \cite{VoiceBox}, or for example the NeMo’s STT En Conformer Transducer Large model, a variant of the model proposed by \cite{Gulati2020}, as used by \cite{SpeechX}.

\subsection{Datasets}
\begin{table}[ht]
\centering
\resizebox{\columnwidth}{!}{
\begin{tabular}{|l|c|c|c|c|}
\hline
\textbf{Paper} &  \textbf{LJSpeech} &\textbf{VCTK} & \textbf{LibriTTS} & \textbf{Additional} \\ \hline
AttentionStitch \cite{AttentionStitch} &  $X$ &$X$ & & \\ \hline  
A3T \cite{A3T} &  $X$ &$X$ & & \\ \hline  
Speechpainter \cite{SpeechPainter} &  &$X$ & $X$ & \\ \hline  
Editts \cite{Editts} &  $X$ && & \\ \hline 
EditSpeech \cite{EditSpeech} &  $X$ &$X$ & & $X^1$ \\ \hline  
FluentSpeech \cite{FluentSpeech} &  &$X$ & $X$ & $X^2$ \\ \hline 
RetrieverTTS \cite{RetrieverTTS} &  && $X$ & \\
Morrison2021 \cite{Morrison2021} &  $X$ && & $X^3$ \\  \hline
\end{tabular}
}
\caption{Comparison of datasets used by the Speech edit community\\
{\tiny 1: MST Aishell-3 (Chinese), 2: SASE, 3: The Everyday Life of Abraham Lincoln read by Bill Boerst}}
\label{tab:datasetsCompare}
\end{table}

\label{dataset_section}
Most Researchers within the speech edit community focus on the three main datasets, as depicted in Table \ref{tab:datasetsCompare}, namely VCTK \cite{VCTK}, a roughly $44h$ newspaper speech dataset, with recordings from $110$ English speakers with various accents. 
LJSpeech \cite{LJSpeech}, a dataset consisting out of 24h short audio clips from a single English speaker as well as LibriTTS \cite{libriTTS}, a roughly $585$ hour dataset without significant background noise.
The issues with those recordings is, that they only consist out of audiobooks, so \cite{VoiceCraft} created REALEDIT, a dataset with  diverse content, accents, speaking styles, recording conditions, and background sounds from audiobooks \cite{libriTTS} YouTube videos \cite{gigaSpeech} and Spotify podcasts \cite{Clifton2020}.

\section{Recent Works}

Recent advancements in speech editing technologies have enabled more precise and natural modifications of audio recordings. These innovations address various challenges such as prosody mismatches and unnatural boundary artifacts, which are common issues in speech editing.

\subsection{VoCo}
The VoCo approach, introduced by \cite{Jin2017}, represents an early method in speech editing that generates a generic voice audio using \gls{tts} technology. This audio is then morphed to match the target voice to be then integrated into the original audio recording.

However, this method often results in unnatural-sounding audio because the generated audio does not match its context. Consequently, this leads to non-smooth boundary artefacts \cite{Morrison2021, AttentionStitch, VoiceCraft, A3T, EditSpeech}.

\subsection{Context-Aware Prosody Correction for Text-Based Speech Editing}
To address those unnatural boundaries within the edit regions due to the prosody mismatches. \cite{Morrison2021} propose using a context-aware method to produce more natural-sounding audio. Their approach utilizes a model based on prior work in neural prosody generation, speech generation, speech manipulation (pitch and time shifting), and speech editing.

First they generate Context-Aware prosody through a series of \gls{nn} from the input phonemes. The networks hereby predict phoneme durations and their pitch.
then they adjust the prosody features by pitch shifting and time stretching, using the TD-PSOLA \cite{TDPSOLA} algorithm. Lastly they remove the artefacts resulting from signal manipulation using a denoising hifi-\gls{gan} ensuring high-fidelity results.

This comprehensive approach enhances the naturalness of edited speech by addressing prosody mismatches and eliminating artifacts, leading to more seamless audio integration.

\subsection{SpeechPainter}

SpeechPainter, developed by \cite{SpeechPainter}, leverages advanced attention mechanisms within its transformer model to seamlessly infill gaps in audio recordings up to one second, maintaining speaker identity, prosody, and recording environment conditions. Additionally, it can generalise to unseen speakers through in-context learning \cite{VoiceCraft, AttentionStitch}.

The model employs a non-\gls{ar} (non-\gls{ar}) architecture based on PerceiverIO \cite{PerceiverIO} to avoid robotic training artefacts in the second training step. It processes data in the form of a (in training randomly) masked, non-overlapping batched log-mel spectrogram, transformed into a $1D$ array and concatenated using $2D$ Fourier positional embeddings. Textual data is padded and transformed into batched $UTF-8$ embeddings with added Fourier positional embeddings and two learned modality embeddings appended to both. These inputs are processed as the key/value in the cross-attention encoder, with the query provided by a learned fixed-size latent.

These values are then fed into $k$ transformer blocks with self-attention. The output of the last self-attention layer is passed to a decoder as key/value, analogous to the encoder, using an output latent. The model then returns a log-mel spectrogram after projecting it to the target spectrogram dimension.

Training was conducted in two phases, both utilising the Adam optimizer. The first phase trains using $L1$ loss on mel-spectrograms to learn correct identification/inpainting of the content of the gap based on the target transcript, maintaining speaker identity, prosody, and the target environment. However, this creates robotic artefacts, which are removed using adversarial training, thus improving perceptual quality. This is achieved by training a discriminator based on \gls{cnn}s to learn the representation of original and created mel-spectrograms. The discriminator is trained with a hinge loss, and the model is compared against it to ensure it is not detected as incorrect by the discriminator.

Experiments were conducted using a neural vocoder\footnote{"Voice Coder" - A synthesizer that produces sounds from an analysis of speech input.} (MelGAN \cite{melGan} architecture) trained on a multi-scale reconstruction and adversarial loss of the wave and stat-based discriminators to synthesize the audio from the log-mel spectrograms. No pre-training or transfer learning was used for the reconstruction of the whole sentence from three-second audio samples. The model utilized two feed-forward neural networks (FFNNs) in the attention blocks.

\subsection{EditSpeech}

Models such as the one proposed by EditSpeech \cite{EditSpeech} condition text generation on the surrounding context to maintain naturalness and quality \cite{A3T, FluentSpeech}. The method proposed by EditSpeech \cite{EditSpeech} divides the audio into edit and non-edit regions. The non-edit regions are directly copied to the output, while the edit regions are processed using an \gls{ar} \gls{ntts} model to generate frames in both forward and backward directions to incorporate context. These generated frames are then merged to create fused mel-spectrograms by selecting the best frames based on context information. This bidirectional fusion ensures smooth boundaries on both sides \cite{EditSpeech, AttentionStitch, VoiceCraft}.

This approach improves upon methods like VoCo \cite{Jin2017}, which rely on unit selection speech synthesis, by enabling the generation of arbitrary text based on context without needing additional audio recordings \cite{EditSpeech}. However, a drawback of this method is its compatibility only with \gls{ar} decoding models and its heavy reliance on speaker embeddings, which limits its transferability to new speakers \cite{A3T}.

The EditSpeech architecture consists of an acoustic model, a duration predictor, and a vocoder. 

The duration predictor forecasts the length of the to-be-inserted sequence based on the edited text. It is trained by minimizing the duration loss between the logs of the ground-truth and the predicted duration. This involves assessing the prediction error and multiplying it by the predicted duration.

The acoustic model employs a text encoder, a speech encoder, a length regulator, a "prenet", and forward and backward decoders. The phone-level text embeddings sequence (obtained by a \gls{g2p} module) is encoded by the encoder. The length regulator expands text embeddings into frame-level embeddings according to the length prediction, aligning the time frames' position. The prenet preprocesses preceding spectrograms and changes their representation to concatenate with the frame-level hidden representation, which is then fed into the decoders. The decoders consist of one forward and one backward decoder, each comprising two unidirectional \gls{lstm}s to synthesize mel-spectrograms. These mel-spectrograms are then stitched together at the point where their frame-level $L2$ norms differ the least.

The vocoder used by EditSpeech is a HiFi-\gls{gan}\cite{hifiGan} \footnote{UNIVERSAL\;1 \url{https://github.com/jik876/hifi-gan}}, which creates speech from a mel-spectrogram.

Training involved several steps for all components in batches using the Adam optimizer. Initially, the data was prepared by pairing speech utterances with text transcriptions and speaker identities. Mel-spectrograms were computed from speech utterances using Tacotron2 \cite{tactron2} signal processing configurations, and ground-truth phone durations were obtained using an \gls{hmm}-based forced aligner.

Training steps began by concatenating the text embeddings, speaker embeddings, and positional embeddings to get the frame-level hidden representation. The mel-spectrograms were synthesized by forward and backward decoders by processing preceding frame representations through a prenet and then through unidirectional \gls{lstm}s, where the decoders minimized both forward and backward mel-spectrogram loss. The duration predictor was trained by minimizing the duration loss between the logs of the ground-truth and predicted durations. Finally, fusion occurred at the point where the frame-level $L2$ norms of the forward and backward predicted mel-spectrograms differed the least, marking the transition from forward to backward generation.

Deletion, insertion, and replacement operations were tested with different presets. For deletion, forced alignment located the start and end times of the phones to remove the speech frames in the input mel-spectrogram. For insertion and replacement, users specified the word position for insertion or text replacement. The \gls{g2p} module processed the original text and replacement text into phone sequences. Forced alignment with the original text determined the duration sequence. The edited phone sequence, duration, and speaker embedding, along with positional embeddings, were then used as input to generate frame-level hidden representations. Forward and backward decoders then synthesized mel-spectrograms, ensuring smooth transitions between edited and non-edited regions.

In general, the system generates mel-spectrograms in partial inference for both forward and backward directions. For unmodified regions, the predicted frame is discarded, and the original frame is fed to the prenet and recurrent decoder for the next frame prediction. For modified regions, the predicted frame is used for subsequent predictions.

\subsection{A3T}
Recently, masked prediction models such as A3T \cite{A3T} have demonstrated improved contextual learning from input mel-spectrograms, significantly enhancing quality and prosody modeling \cite{FluentSpeech}. 

A3T achieves this by utilizing forced alignment and \gls{g2p} conversion, similar to EditSpeech \cite{EditSpeech}, using the HTK \cite{HTK} method and an external English duration predictor based on FastSpeech2 \cite{fastspeech2}. The duration predictor, analogous to the one in EditSpeech, predicts the length of the masked segments, taking into account the ratio of original to edited segments. This ensures the alignment and timing of the generated spectrogram are coherent with the original input. These steps in combination ensure accurate phoneme alignment and duration prediction.

The model uses cross-modal alignment embeddings in its BERT-style architecture. By combining position and alignment embeddings on top of the text and speech embeddings, which are then concatenated, the model aligns text and audio more effectively. This approach is an enhancement over EditSpeech's method of simply concatenating text and speaker embeddings with positional embeddings, as it enables the prediction of contextually coherent speech using masked reconstruction with the Conformer \cite{conformer} architecture \cite{VoiceCraft, AttentionStitch}. 
 
A3T operates by using these embeddings in combination with a non-\gls{ar} encoder to predict high-quality spectrograms, which are then refined using a five-layer \gls{cnn} post-net. This refinement step helps generate more accurate and high-quality spectrograms, a step not present in the EditSpeech model. Additionally, it allows for faster inference and avoids some of the pitfalls of \gls{ar} models, such as error accumulation.

The encoder and decoder are realized using the Conformer architecture, which integrates a \gls{cnn} module and a feed-forward module. In experiments, A3T performed better than the traditional transformer in acoustic text processing, indicating an improvement over \cite{EditSpeech}'s \gls{lstm}-based forward and backward decoders.

Finally, the spectrogram is converted into an audio wave using a Parallel Wave\gls{gan} \cite{waveGAN} vocoder, different from \cite{EditSpeech}'s use of HiFi-\gls{gan}.

A3T is trained by masking random utterances and phones in the middle of sentences during training. Similar to the insertion and replacement process by \cite{EditSpeech}, the embeddings are passed through eight Conformer layers. The output of these layers is combined with the post-net output to generate the final mel-spectrogram. The model is evaluated on its ability to reconstruct masked speech segments (repainting) conditioned on ground truth. The repainted speech is compared to human re-recordings to assess quality and naturalness.

A downside of this approach is that it assumes a deterministic mapping from text and context to the target, which is only realistic for very short segments \cite{VoiceBox}.

\subsection{VoiceBox}
 \cite{VoiceBox} proposed VoiceBox, a generative model that uses non-\gls{ar} flow matching to infill speech given audio context and text. It can perform several tasks such as noise removal (through infilling), content editing, and style conversion \cite{Mapache, VoiceCraft, VoiceBox, Editts}. Unlike models such as A3T from \cite{A3T} and SpeechPainter \cite{SpeechPainter}, which are only feasible for short segments and assume a deterministic mapping from text and context to the target, \cite{VoiceBox}'s approach can infill data of any length and does not assume such a deterministic mapping. Additionally, VoiceBox can be trained on "in-the-wild" datasets with significant variation.

The infilling approach of VoiceBox involves generating speech segments based on surrounding audio context and corresponding text transcripts. Unlike previous models, VoiceBox does not rely on explicit audio style labels (e.g., speaker identity or emotion), making it more scalable and adaptable to diverse generative tasks. \cite{VoiceBox}

VoiceBox is built on a non-\gls{ar} continuous normalizing flow (CNF) architecture. It trains the CNF model to map a simple prior distribution to the target speech distribution and uses in-context learning by utilizing both preceding and succeeding context during generation, making it suitable for editing tasks where only a segment of speech needs to be generated. This contrasts with \gls{ar} models, which generate speech sequentially from start to end.

Lastly, VoiceBox decouples audio and duration modeling, allowing for more precise control over alignment and the integration of continuous features. Although A3T and EditSpeech incorporate duration predictors, VoiceBox's approach offers more flexibility in handling different generative tasks.

\subsection{SpeechX}
The model proposed in \cite{SpeechX} combines neural codec language modeling with a non-\gls{ar} transformer to generate codec or acoustic tokens based on input, employing multi-task learning with task-dependent prompting \cite{Mapache}. Unlike \cite{A3T}, which is restricted to clean signals and lacks the ability to modify spoken content while preserving background sounds, this approach addresses these limitations for "noisy speech editing". Additionally, unlike VoiceBox, which expects noisy signals to be surrounded by clean segments, this method does not have such constraints due to its design that allows task prompt tasks. This model preserves background noise during speech editing \cite{SpeechX}. However, its performance is constrained by the accuracy of the neural codec model for acoustic tokenization \cite{SpeechX}.

SpeechX integrates several advanced components to achieve its goals, distinguishing itself from previous models such as SpeechPainter, EditSpeech, A3T, and VoiceBox. SpeechX employs EnCodec \cite{encodec}, a neural codec modeling architecture (encoder-decoder), producing a sequence of neural codecs that are converted to waveforms using a codec decoder. It receives a \gls{g2p}, like A3T and EditSpeech, for text input for semantic tokens, and audio tokens are obtained by a neural codec model encoder. Both are embedded using sinusoidal style embedding projection.

The model uses both non-\gls{ar} and \gls{ar} transformer models. The \gls{ar} model outputs neural codes, while the non-\gls{ar} model generates neural codes for all layers, balancing generation flexibility and inference speed. During training, SpeechX was trained by swapping randomly between tasks like noise suppression and speech removal to prevent overfitting.

\subsection{FluentSpeech}
FluentSpeech \cite{FluentSpeech} is a context-aware, diffusion-based speech editing model designed for stutter removal without the over-smoothing seen in previous approaches \cite{VoiceCraft}. It iteratively refines modified mel-spectrograms with guidance from context features and a stutter predictor injected into its hidden sequence.

Building upon the works of Tan, Jin, Wang, and Bai \cite{EditSpeech, Jin2017, Wang2022, A3T}, FluentSpeech aims to learn better contextual information from input mel-spectrograms. FluentSpeech outperforms these methods in terms of quality and prosody modeling. Previous methods only addressed reading-style speeches, leaving stutter removal a significant challenge \cite{FluentSpeech}. Traditional stutter removal techniques often result in blurry mel-spectrograms lacking detail, creating unnatural sounds in the edited regions and requiring manual determination of the utter regions. FluentSpeech addresses these issues automatically using its stutter predictor \cite{FluentSpeech}, which localizes stutter regions and injects stutter embeddings into the text's hidden sequence to reduce discrepancies between the text and the stuttering speech recording \cite{FluentSpeech}.

However, it is important to note that FluentSpeech was only tested on English speech data, which may limit its generalizability to other languages and dialects \cite{FluentSpeech}.

FluentSpeech's architecture integrates a linguistic encoder, which is transformer-based, transforming the phoneme sequence of the transcription into a hidden text sequence. This approach aligns with the transformer architecture used in A3T but focuses specifically on stutter removal. A context-aware spectrogram denoiser in the form of a WaveNet \cite{waveNet} aggregates various features, including phoneme embeddings, acoustic embeddings, pitch embeddings, and stutter embeddings, similar to the context conditioning module in EditSpeech. This enhances the quality of the reconstructed mel-spectrograms, in conjunction with the stutter predictor.

Due to the modality gap between text and speech, FluentSpeech emphasizes alignment modeling \cite{FluentSpeech}. It employs a masked duration predictor to achieve fluent duration transitions in edited regions.

\subsection{VoiceCraft}
VoiceCraft \cite{VoiceCraft} introduces a regressive transformer decoder architecture for neural codec token infilling, similar to SpeechX, but with a distinct emphasis on \gls{ar} prediction and bidirectional context. Their model refines \gls{ar} sequence prediction and uses masked delayed stacking to enable generation within an existing sequence. VoiceCraft rearranges output tokens through causal masking with \gls{ar} continuation or infilling within the bidirectional context and employs two delayed stacking mechanisms (multi-code block modeling).

The causal masking procedure moves the masked spans to the end of the sequence, allowing the model to use both past and future unmasked tokens during infilling. This bidirectional context is a significant advancement over the unidirectional approaches in models like EditSpeech. The delayed stacking mechanism ensures efficient multi-codebook modeling by delaying the integration of different codebook tokens until a later stage in the process.

The utilization of delayed stacking and causal masking contrasts with the cross-modal alignment embeddings in A3T and the forward-backward decoding in EditSpeech. This allows VoiceCraft to achieve more accurate and contextually coherent infilling and continuation compared to the unidirectional approaches in A3T and EditSpeech.

The model employs a transformer decoder architecture, trained on \gls{ar} sequence prediction. During training, spans of tokens are randomly masked, and the model predicts these masked tokens based on the unmasked tokens, similar to the masked reconstruction approach in A3T but with different masking and stacking techniques. VoiceCraft addresses limitations seen in VoiceBox, which expects noisy signals to be surrounded by clean segments. VoiceCraft's flexible approach to sequence generation makes it more adaptable to various audio conditions without such constraints. However, it still faces some limitations, such as occasional long silences or scratching sounds. \cite{VoiceCraft}
 
\subsection{Attention Stitch}
AttentionStitch \cite{AttentionStitch} separates the \gls{tts} generation from the actual editing process, morphing the \gls{tts}-generated data into the edited portion. This is achieved by using a pre-trained \gls{tts} model and incorporating a double attention block network on top of it to automatically merge the synthesized mel-spectrogram with that of the edited text. According to their measurements, this approach enhances the naturalness of the stitched audio by creating a smoother audio segment.

Unlike SpeechPainter, EditSpeech, A3T, and FluentSpeech, which integrate the generation and editing processes, AttentionStitch separates \gls{tts} generation from the editing process. This separation allows for more precise control over the editing process and the use of pre-trained high-quality \gls{tts} models, leading to a unique model architecture.

The model uses a pre-trained FastSpeech2 \cite{fastspeech2} \gls{tts} model similar to \cite{A3T}, taking phonemes of the edited text as input to generate an initial mel-spectrogram. This contrasts with the models in EditSpeech, VoiceBox, and SpeechX, which do not explicitly utilize pre-trained \gls{tts} models for this purpose. This synthesized mel-spectrogram is then concatenated with a masked reference mel-spectrogram (with $10\%$ of its content near the center masked). On top of this, a double attention block network is used to merge the synthesized mel-spectrogram with the edited text's mel-spectrogram, ensuring a smooth transition between the synthesized and reference segments.

This approach is distinct from the cross-modal alignment embeddings used in A3T, the forward-backward decoding in EditSpeech, and the diffusion-based approach in FluentSpeech. To refine the output, AttentionStitch uses a post-net module and a HiFi-\gls{gan} vocoder similar to the approach of \cite{Morrison2021} to transform the mel-spectrogram into a waveform. Skip connections are employed between the output of the double attention block and the post-net, as they have proven beneficial in speech synthesis \cite{AttentionStitch, 7953221, 8706591}.

During training, masking is performed by replacing corresponding parts of the mel-spectrogram with zeros, and the reference text is modified by replacing words with target words. To ensure the reference mel-spectrogram has the same length, they use the duration predictor of FastSpeech2 to resize the mask accordingly, similar to the approach used in A3T. The speech editing operation takes place within the double attention block, where a mean average error loss is used for both the double attention block and the post-net.

AttentionStitch deliberately avoids comparing with fully resynthesized edited text because the final audio sample, though indistinguishable in the edited part, differs from the reference audio. For the VCTK dataset, AttentionStitch is compared to competitive methods like A3T and EditSpeech. Models like SpeechPainter were excluded from the comparison due to their specific limitations and evaluation contexts, as SpeechPainter is limited to filling small gaps with the same text.

\subsection{Mapache}
 
\gls{ar} modeling, which limits speech synthesis to left-to-right generation, is unsuitable for producing speech edits free from audio discontinuities \cite{Mapache}. Mapache addresses this limitation by utilizing a non-\gls{ar} architecture with parallel sequence-to-sequence transformers to model discrete text and speech representations, allowing for both speech editing and synthesis \cite{Mapache}. Unlike SpeechX, EditSpeech, and Voicecraft that employ \gls{ar} models, this approach enables more flexible and natural speech editing even if there are discontinuities in the audio, which cause difficulties to other \gls{ar} models.

The architecture of Mapache consists of four main components. First, a speech tokenizer uses a VQ-VAE (Vector Quantized Variational Autoencoder) \cite{vqvae} with an RNN prosody encoder post VQ codebook to convert the waveform into tokenized log-mel spectrograms. These spectrograms are then tokenized using the VQ encoder, with optimization based on mean squared error and commitment loss between the predicted and actual spectrogram.

The core component of Mapache is the parallel language modeling transformer, speechMPT, which takes the tokenized speech and text inputs to create embeddings and speech tokens. The transformer is composed of $30$ blocks, each consisting of two $16$-head attention layers followed by a feed-forward layer (using GeGLU \cite{geGLU}) with skip connections \cite{Mapache}. The first attention layer performs self-attention over the masked and token embeddings, while the second layer performs cross-attention between text and speech embeddings, similar to the mechanisms used by \cite{SpeechPainter}.

Next, a diffusion decoder (DiT-DDPM\footnote{based on DDPM \url{https://github.com/openai/improved- diffusion} and \cite{dit}}) upscales the speech tokens or embeddings into log-mel spectrograms. This component uses denoising diffusion probabilistic models (DDPMs) and incorporates self-attention layers to refine the embeddings, enhancing speaker identity and speech articulation in the output. This is similar to the approaches used by \cite{FluentSpeech} and \cite{VoiceBox}, which also incorporate denoising and refinement techniques to enhance the quality of the generated speech.

Finally, a vocoder (UnivNet \cite{univNet}) synthesizes the log-mel spectrograms into waveforms, completing the process.

\section{Discussion}

The comparison between the models is not straightforward due to several factors: some models are not openly accessible like \cite{SpeechX, SpeechPainter}, are trained on partially inaccessible files like \cite{Mapache}, or are trained on a mixture of datasets that differ in quality and focus like  \cite{Mapache, FluentSpeech, SpeechX, VoiceCraft}.
Moreover, even if they were trained on the same dataset, their method of how they edited the data, prepossessed it as well as tested it on which task with which method differs so greatly, that it is almost impossible to compare two publications directly. 

Additionally, the term "speech editing" is loosely defined, encompassing even smaller sub-parts of the process, and papers often use qualitative or different incomparable measures, making direct comparisons difficult. However, we attempt to address these issues in this section.

As stated previously, even though the datasets are different and have varying focuses, as discussed in Section \ref{dataset_section} and highlighted in Figure \ref{tab:datasetsCompare}, most models are tested on the VCTK \cite{VCTK} and LJSpeech \cite{LJSpeech} datasets. This commonality allows us to group them together and examine their results. We have grouped the objective measures in Table \ref{table:objective_models_comparison} and the subjective measures in Table \ref{table:subjective_models_comparison}, organized by their dataset and sorted by their measure.

As shown in Table \ref{table:objective_models_comparison}, most models use the \gls{mos} to evaluate their performance. However, the \gls{mos} measurements themselves vary across research papers. Some studies measure the overall quality of speech \cite{VoiceBox, FluentSpeech}, while others break it down into subcategories like naturalness, intelligibility, and speaker similarity \cite{VoiceCraft, FluentSpeech}, or do not evaluate it distinctively against other baselines \cite{A3T}. Moreover, an intrinsic problem with \gls{mos} is that different raters might weigh its components differently, making it difficult to compare across papers as described in Section \ref{mos-section}.

While it is feasible to compare different models within a single paper using the \gls{mos} measure, comparisons between papers are not possible due to its subjective nature. Therefore, we also created a table showing which papers have been compared using which measure, as seen in Table \ref{table:paper_comparison}.

Another subjective measure used by \cite{Mapache} is the \gls{mushra}. \gls{mushra} is a measure approved by the \gls{itu}\footnote{https://www.itu.int/rec/R-REC-BS.1534-3-201510-I/en} as a standard method for the subjective assessment of intermediate quality levels of audio systems. However, apart from \cite{Mapache}, no other featured paper here used it to measure subjective performance. The scores for the subjective measures can be found in Table \ref{table:objective_models_comparison}.

The research papers reviewed also include objective measures, as illustrated in Table \ref{table:subjective_models_comparison}. Most research teams used \gls{wer} and similarity measures, which are common for \gls{tts} tasks. This is because \gls{tts} is a significant component of speech editing, as newer papers often first generate text using \gls{tts} and then morph it into the context speaker’s voice \cite{Morrison2021, AttentionStitch, VoiceCraft, A3T, EditSpeech}. Nevertheless, there are noticeable differences, such as the use of objective similarity measures "sim-o" and "sim-r," which differ from other similarity measures. Apart from \gls{wer} and similarity measures, other measures such as \gls{pesq} \cite{Rix2001}, \gls{stoi} \cite{STOI}, and \gls{mcd} \cite{mcd} are used. \gls{pesq} and \gls{stoi} indicate speech quality and intelligibility, respectively \cite{FluentSpeech}, and were employed by EditSpeech \cite{EditSpeech} to compare models. However, the biggest challenge is that due to varying task descriptions—like inpainting of a word, phrase, or syllable, and noisy versus clean speech editing—papers are difficult to compare directly.

\begin{table}[ht]
\centering
\resizebox{\columnwidth}{!}{

  \begin{tabular}{|l|l|l|l|}
    \hline
        \textbf{Paper} & \textbf{Language} & \textbf{Multi speaker} & \textbf{Unique Attribute}\\ \hline
        VOCO \cite{Jin2017}& English & Yes & ~ \\ 
        AttentionStich \cite{AttentionStitch} & English & Yes & ~ \\ 
        VoiceCraft \cite{VoiceCraft} & English & Yes & {\tiny Realistic, Noisy data} \\ 
        Mapache \cite{Mapache}& English & Yes & {\tiny Amazon inhouse, Can not be evaluated} \\ 
        Voicebox \cite{VoiceBox}& English\textsuperscript{1} & Yes & {\tiny Realistic, Noisy data} \\ 
        A3T \cite{A3T}& English & Yes & ~ \\ 
        Speechpainter \cite{SpeechPainter}& English & Yes & ~ \\ 
        Editts \cite{Editts}& English & No & ~ \\ 
        Editspeech \cite{EditSpeech}& English\textsuperscript{2} & Yes & ~ \\ 
        Fluentspeech \cite{FluentSpeech}& English & Yes & ~ \\
        SpeechX \cite{SpeechX}& English & Yes & {\tiny Noisy data} \\ 
        RetrieverTTS \cite{RetrieverTTS}& English & Yes & ~ \\ 
        Morrison \cite{Morrison2021} & English & No & ~\\\hline
    \end{tabular}

    }
\caption{Depiction of the languages the models were trained and tested with \\ 
    {\tiny $1$: it was additionaly also evaluated on French, German, Spanish, Polish, and Portuguese; $2$: Chinese}}
\label{table:data_languages}
\end{table}

Some models, such as \cite{SpeechPainter}, were not tested on any objective measures, or were incomparable due to different focuses and single-versus-multi-speaker scenarios, as highlighted in Table \ref{table:data_languages}. While some research teams emphasized the use of many languages and tested their models accordingly, such as \cite{EditSpeech} with Chinese and \cite{VoiceBox}, which tested without multiple languages, others emphasized multi-speaker datasets with multiple English accents for robustness. Only the works of \cite{Editts} and \cite{Morrison2021} focused on single-speaker text editing using the LJSpeech \cite{LJSpeech} dataset.

An interesting finding for zero-shot performance is that newer models place more emphasis on the use of realistic, in-the-wild data to enhance their models with data that has worse acoustic quality, noise, or stuttering \cite{VoiceCraft, Mapache, VoiceBox}. In some cases, data was deliberately mixed with noisy samples to degrade the quality of the training and test datasets \cite{SpeechX}.

\section{Conclusion}
Speech editing is an established discipline that has recently been significantly advanced by the rise of Transformers \cite{Vaswani2017} and their corresponding attention layers. This technological leap has brought new capabilities and precision to the field. Moreover, the growing interest from major tech companies in speech editing is likely to accelerate the development of advanced speech editing systems.

Transformers, such as ChatGPT and other models, have already integrated into everyday life, and the next major development is likely to be enhanced \gls{tts} systems, which play a crucial role in speech editing. Our contribution with this paper aims to leverage current knowledge and inspire more researchers and students to delve into the topic of speech editing. By providing a comprehensive comparison of the latest methods, their similarities, challenges, and drawbacks, we hope to offer valuable insights and foster increased interest and innovation in this field.

\label{sec:bibtex}

\bibliography{acl_latex}

\appendix

\section{Appendix}
\label{sec:appendix}

\begin{table}[ht]
\centering
\resizebox{\columnwidth}{!}{

    \begin{tabular}{ccccccccccc}
    \hline
        Task & Measure & Mapache & A3t & Attention Stich & Edit speech & FluentSpeech & VoiceCraft & SpeechX (Vall-E) & VoiceBox & Speechpainter \\ \hline
        Testset: Custom VoiceBox & ~ & ~ & ~ & ~ & ~ & ~ & ~ & ~ & ~ & ~ \\ \hline
        Noise Removal & SIM ↑-o & ~ & 0.148 & ~ & ~ & ~ & ~ & ~ & 0.612 & ~ \\ 
        Zero shot TTS cross sentence & SIM ↑-o & ~ & 0.046 & ~ & ~ & ~ & ~ & ~ & 0.662 & ~ \\ 
        Zero shot TTS cross sentence & SIM ↑-r & ~ & 0.146 & ~ & ~ & ~ & ~ & ~ & 0.681 & ~ \\ 
        Noise Removal & WER ↓ & ~ & 11.5 & ~ & ~ & ~ & ~ & ~ & 2 & ~ \\ 
        Zero shot TTS cross sentence & WER ↓ & ~ & 63.3 & ~ & ~ & ~ & ~ & ~ & 1.9 & ~ \\ \hline
        Testset: LibriLight & ~ & ~ & ~ & ~ & ~ & ~ & ~ & ~ & ~ & ~ \\ \hline
        Clean Speech editing & SIM ↑ & ~ & 0.29 & ~ & ~ & ~ & ~ & 0.76 & ~ & ~ \\ 
        Noisy Speech editing & SIM ↑ & ~ & 0.18 & ~ & ~ & ~ & ~ & 0.65 & ~ & ~ \\ 
        Clean Speech editin & WER ↓ & ~ & 17.17 & ~ & ~ & ~ & ~ & 5.63 & ~ & ~ \\ 
        Noisy Speech editing & WER ↓ & ~ & 32.17 & ~ & ~ & ~ & ~ & 13.95 & ~ & ~ \\ \hline
        Testset: LibriTTS & ~ & ~ & ~ & ~ & ~ & ~ & ~ & ~ & ~ & ~ \\ \hline
        Speech Editing & MCD ↓ & ~ & 6.25 & ~ & 6.92 & 5.86 & ~ & ~ & ~ & ~ \\ 
        Speech Editing & PRESQ ↑ & ~ & 1.18 & ~ & 1.43 & 1.91 & ~ & ~ & ~ & ~ \\ 
        Zero shot tts & SIM ↑ & ~ & ~ & ~ & ~ & 0.47 & 0.55 & ~ & ~ & ~ \\ 
        Speech Editing & STOI ↑ & ~ & 0.41 & ~ & 0.69 & 0.81 & ~ & ~ & ~ & ~ \\ 
        Zero shot tts & WER ↓ & ~ & ~ & ~ & ~ & 3.5 & 4.5 & ~ & ~ & ~ \\ \hline
        Testset: Mapache Superset & ~ & ~ & ~ & ~ & ~ & ~ & ~ & ~ & ~ & ~ \\ \hline
        Inpainting (phrase) & SIM ↑ & 0.93 & 0.74 & ~ & ~ & ~ & ~ & ~ & ~ & ~ \\ 
        Inpainting (Syllable) & SIM ↑ & 0.95 & 0.9 & ~ & ~ & ~ & ~ & ~ & ~ & ~ \\ 
        Inpainting (word) & SIM ↑ & 0.94 & 0.85 & ~ & ~ & ~ & ~ & ~ & ~ & ~ \\ 
        Inpainting (phrase) & WER ↓ & 4 & 21 & ~ & ~ & ~ & ~ & ~ & ~ & ~ \\ 
        Inpainting (Syllable) & WER ↓ & 5.5 & 9.8 & ~ & ~ & ~ & ~ & ~ & ~ & ~ \\ 
        Inpainting (word) & WER ↓ & 4.1 & 12.3 & ~ & ~ & ~ & ~ & ~ & ~ & ~ \\ \hline
        Testset: RealEddit & ~ & ~ & ~ & ~ & ~ & ~ & ~ & ~ & ~ & ~ \\ \hline
        Speech Editing & WER ↓ & ~ & ~ & ~ & ~ & 4.5 & 6.1 & ~ & ~ & ~ \\ \hline
        Testset: VCTK & ~ & ~ & ~ & ~ & ~ & ~ & ~ & ~ & ~ & ~ \\ \hline
        Speech Editing & MCD ↓ & ~ & 5.69 & ~ & 5.33 & 4.74 & ~ & ~ & ~ & ~ \\ 
        Text edit & MCD ↓ & ~ & 7.97 & 6.5 & 7.54 & ~ & ~ & ~ & ~ & ~ \\ 
        Speech Editing & PRESQ ↑ & ~ & 1.39 & ~ & 1.35 & 1.82 & ~ & ~ & ~ & ~ \\ 
        Speech Editing & STOI ↑ & ~ & 0.7 & ~ & 0.68 & 0.78 & ~ & ~ & ~ & ~ \\ \hline
    \end{tabular}

    }
\caption{Comparison of the \textbf{objective} measures and results from the presented papers for their respective tasks on various test sets and metrics.\\
{\tiny The ↑,↓ arrows mark, that higher and lower values are preferred respectively.}} 
\label{table:objective_models_comparison}
\end{table}
\newpage
\begin{table}[ht]
\centering
\resizebox{\columnwidth}{!}{

    \begin{tabular}{ccccccccccc}
    \hline
        Task & Measure & Mapache & A3t & Attention Stich & Edit speech & FluentSpeech & VoiceCraft & SpeechX (Vall-E) & VoiceBox & Speechpainter \\ \hline
        Testset: Custom VoiceBox & ~ & ~ & ~ & ~ & ~ & ~ & ~ & ~ & ~ & ~ \\ \hline
        Noise Removal & MOS ↑(Speech Quality) & ~ & 3.10±0.15 & ~ & ~ & ~ & ~ & ~ & 3.87±0.17 & ~ \\ \hline
        Testset: LibriTTS & ~ & ~ & ~ & ~ & ~ & ~ & ~ & ~ & ~ & ~ \\ \hline
        SpeechEditing & MOS ↑(Intelligibility) & ~ & ~ & ~ & ~ & 3.89±0.09 & 4.05±0.08 & ~ & ~ & ~ \\ 
        SpeechEditing & MOS ↑(naturalness) & ~ & ~ & ~ & ~ & 3.81±0.06 & 4.03±0.05 & ~ & ~ & ~ \\ 
        Zero shot tts & MOS ↑Intelligibility & ~ & ~ & ~ & ~ & 3.7±0.11 & 4.38±0.08 & ~ & ~ & ~ \\ 
        Zero shot tts & MOS ↑Naturalness & ~ & ~ & ~ & ~ & 3.34±0.11 & 4.16±0.08 & ~ & ~ & ~ \\ 
        Zero shot tts & MOS ↑Speaker Similarity & ~ & ~ & ~ & ~ & 4.10±0.09 & 4.35±0.08 & ~ & ~ & ~ \\ \hline
        Testset: LibriTTS Clean & ~ & ~ & ~ & ~ & ~ & ~ & ~ & ~ & ~ & ~ \\ \hline
        Inpainting (<1s) & MOS ↑(Naturalness) & ~ & ~ & ~ & ~ & ~ & ~ & ~ & ~ & 3.60 ± 0.09 \\ \hline
        Testset: LibriTTS Other & ~ & ~ & ~ & ~ & ~ & ~ & ~ & ~ & ~ & ~ \\ \hline
        Inpainting (<1s) & MOS ↑(Naturalness) & ~ & ~ & ~ & ~ & ~ & ~ & ~ & ~ & 3.12 ± 0.10 \\ \hline
        Testset: Mapache Superset & ~ & ~ & ~ & ~ & ~ & ~ & ~ & ~ & ~ & ~ \\ \hline
        Inpainting (phrase) & MUSHRA ↑ & 76.1 & 45.7 & ~ & ~ & ~ & ~ & ~ & ~ & ~ \\ 
        Inpainting (Syllable) & MUSHRA ↑ & 76.3 & 62.1 & ~ & ~ & ~ & ~ & ~ & ~ & ~ \\ 
        Inpainting (word) & MUSHRA ↑ & 77.3 & 55.9 & ~ & ~ & ~ & ~ & ~ & ~ & ~ \\ \hline
        Testset: RealEddit & ~ & ~ & ~ & ~ & ~ & ~ & ~ & ~ & ~ & ~ \\ \hline
        Speech Editing & MOS ↑(Intelligibility) & ~ & ~ & ~ & ~ & 3.97±0.05 & 4.11±0.05 & ~ & ~ & ~ \\ 
        Speech Editing & MOS ↑(naturalness) & ~ & ~ & ~ & ~ & 3.81±0.06 & 4.03±0.05 & ~ & ~ & ~ \\ \hline
        Testset: VCTK & ~ & ~ & ~ & ~ & ~ & ~ & ~ & ~ & ~ & ~ \\ \hline
        Speech editing (Insert) & MOS ↑ & ~ & 3.53±0.17 & ~ & 3.50±0.16 & ~ & ~ & ~ & ~ & ~ \\ 
        Speech editing (Replace) & MOS ↑ & ~ & 3.65±0.15 & ~ & 3.58±0.16 & ~ & ~ & ~ & ~ & ~ \\ 
        Text edit & MOS ↑ & ~ & 3.3±0.35 & 3.51±0.23 & 3.28±0.33 & ~ & ~ & ~ & ~ & ~ \\ 
        Inpainting (<1s) & MOS ↑(Naturalness) & ~ & ~ & ~ & ~ & ~ & ~ & ~ & ~ & 3.70 ± 0.09 \\ 
        Speech Editing & MOS ↑(Speaker Similarity Seen & ~ & 4.27±0.09 & ~ & 4.26±0.10 & 4.42±0.06 & ~ & ~ & ~ & ~ \\ 
        Speech Editing & MOS ↑(Speaker Similarity Unseen & ~ & 3.53±0.14 & ~ & 3.90±0.13 & 4.21±0.11 & ~ & ~ & ~ & ~ \\ 
        Speech Editing (Seen speaker) & MOS ↑(Speech Quality) & ~ & 4.09±0.10 & ~ & 4.00±0.10 & 4.27±0.11 & ~ & ~ & ~ & ~ \\ 
        Speech Editing (unseen speaker) & MOS ↑(Speech Quality) & ~ & 3.90±0.10 & ~ & 3.89±0.09 & 4.18±0.09 & ~ & ~ & ~ & ~ \\ \hline
    \end{tabular}

}
\caption{Comparison of the \textbf{subjective} measures and results from the presented papers for their respective tasks on various test sets and metrics.\\
{\tiny The ↑,↓ arrows mark, that higher and lower values are preferred respectively.}}
\label{table:subjective_models_comparison}
\end{table}

\begin{table}[ht]
\centering
\resizebox{\columnwidth}{!}{

    \begin{tabular}{ccccccc}
    \hline
        ~ & EditSpeech 2021 & A3T & VoiceBox & FluentSpeech & Editts & SpeechPainter \\ \hline
        EditSpeech 2021 & ~ & ~ & ~ & ~ & ~ & ~ \\  \hline
        A3T & \begin{tabular}[c]{@{}c@{}}{\tiny \textbf{Tasks:} Identity speech reconstruction}\\{\tiny \textbf{Using:} \gls{mcd}}\end{tabular} & ~ & ~ & ~ & ~ & ~ \\ \hline
        VoiceBox & ~ & \begin{tabular}[c]{@{}c@{}}{\tiny \textbf{Tasks:} Denoising, zero shot \gls{tts}}\\{\tiny \textbf{Using:} \gls{wer}, \gls{sim}, \gls{qmos}}\end{tabular}& ~ & ~ & ~ & ~ \\ \hline
        SpechPainter & ~ & ~ & ~ & ~ & ~ & ~ \\ \hline
        SpeechX & ~ & \begin{tabular}[c]{@{}c@{}}{\tiny \textbf{Tasks:} Speech Edit}\\{\tiny \textbf{Using:} \gls{wer}, \gls{sim}}\end{tabular}& ~ & ~ & ~ & ~ \\ \hline
        FluentSpeech & \begin{tabular}[c]{@{}c@{}}{\tiny \textbf{Tasks:} Speech Edit (speech quality, speaker similarity, intelligibility)}\\
        {\tiny \textbf{Using:} \gls{mos}, \gls{mcd}, \gls{stoi}, \gls{pesq}}\end{tabular} & \begin{tabular}[c]{@{}c@{}}{\tiny \textbf{Tasks:} Speech Edit (speech quality, speaker similarity, intelligibility)}\\
        {\tiny \textbf{Using:} \gls{mos}, \gls{mcd}, \gls{stoi}, \gls{pesq}}\end{tabular}  & ~ & ~ \\ \hline
        VoiceCraft & ~ & ~ & ~ & \begin{tabular}[c]{@{}c@{}}{\tiny \textbf{Tasks:} Speech Edit (Intelligibility, Naturalness)}\\
        {\tiny \textbf{Using:} \gls{mos}}\end{tabular}  & ~ \\ \hline
        AttentionStitch & \begin{tabular}[c]{@{}c@{}}{\tiny \textbf{Tasks:} Inpainting}\\
        {\tiny \textbf{Using:} \gls{mos}, \gls{mcd}}\end{tabular} & \begin{tabular}[c]{@{}c@{}}{\tiny \textbf{Tasks:} Inpainting}\\
        {\tiny \textbf{Using:} \gls{mos}, \gls{mcd}}\end{tabular} & ~ & ~ & {\tiny no\textsuperscript{*}}  & {\tiny no\textsuperscript{*}} \\ \hline
        Mapache & ~ & \begin{tabular}[c]{@{}c@{}}{\tiny \textbf{Tasks:} Inpainting, Repainting (words, sentences)}\\
        {\tiny \textbf{Using:} \gls{wer}, \gls{sim}, \gls{mushra}}\end{tabular} & ~ & ~ & ~ & ~ \\ \hline
    \end{tabular}

    }
\caption{The paper on the Y axis (sorted from the oldest to newest) were compared with the paper on the X axis\\{\tiny \textsuperscript{*}: based on the nature of the works, \cite{AttentionStitch} didn't want to be compared with them}}
\label{table:paper_comparison}
\end{table}

\begin{table}[ht]
\centering
\resizebox{\columnwidth}{!}{

  \begin{tabular}{|l|l|}
    \hline
        \textbf{Paper}& \textbf{Model Type}\\ \hline
        VOCO \cite{Jin2017}& non-\gls{ar}\\ 
        Morrison \cite{Morrison2021} & non-\gls{ar}\\
        Speechpainter \cite{SpeechPainter}& non-\gls{ar}\\ 
        Editspeech \cite{EditSpeech}& \gls{ar}\\ 
        A3T \cite{A3T}& \gls{ar}\\ 
        Voicebox \cite{VoiceBox}& non-\gls{ar}\\ 
        SpeechX \cite{SpeechX}& non-\gls{ar}\\ 
        Fluentspeech \cite{FluentSpeech}& non-\gls{ar}\\
        VoiceCraft \cite{VoiceCraft} &  \gls{ar}\\ 
        AttentionStich \cite{AttentionStitch} & non-\gls{ar} \\ 
        Mapache \cite{Mapache}& non-\gls{ar}\\ 
        
        \hline
    \end{tabular}

    }
\caption{Depiction of the languages the models were trained and tested with \\ 
    {\tiny $1$: it was additionally also evaluated on French, German, Spanish, Polish, and Portuguese; $2$: Chinese}}
\label{table:ar_nar_models}
\end{table}

\end{document}